\magnification=\magstep1
\voffset=0.5 true cm
\hoffset=0.66cm
\hsize=15.0 true cm
\baselineskip=12pt
\frenchspacing
\parindent=13pt
\headline={\ifnum\pageno=1\hss\else\hss - \folio\ - \hss\fi}
\footline={\hss}
\font\meinfont=cmbx10 scaled \magstep1
\noindent{\meinfont Comment on testing Bell's inequality with Rydberg atoms entangled through cavity photons}
\par
\medskip\bigskip
\noindent {\bf Arthur Jabs}
\medskip
\par
\noindent Physics Department, Federal University of Para\'\i ba, 58.059-970 Jo\~ao Pessoa, Brazil
\par
\noindent (E-mail: drajabs@aol.com)
\par
\medskip
\noindent
(November 1998)
\par
\bigskip \medskip

\noindent {\bf Abstract.} Rydberg atoms traversing a micromaser cavity one after the other can emerge in correlated states, and according to standard Copenhagen quantum mechanics this may lead to a violation of Bell's inequality, that is, to a Bell sum $S \!> \! 2$. {\it S} is here calculated in various ideal physical situations including those with (1) different initial states of the atoms, (2) different free adjustable parameters, being either the phase or the turning angle of a classical radiation field put behind the cavity, and (3) either equal or different Rabi angles for successive atoms. The experiments are crucial for a realist interpretation which predicts $S \! \le \! 2$ because particles of different kinds (photons and atoms) are involved.
\medskip\bigskip
\par
\hrule
\def \state#1{\vert #1 \rangle}
\def\footnoterule{\kern-3pt
\hrule width 2 true in
\kern 2.6 pt}
\footnote{}{This paper is also available via internet:
\par\noindent 
http://xxx.lanl.gov/abs/quant-ph/9811042
\par\noindent
or e-mail: quant-ph@lanl.gov, subject: get 9811042}
\medskip\bigskip\par

\noindent
{\bf 1~~Introduction}
\smallskip\par\noindent
Up to the present time violations of Bell's inequality have been demonstrated experimentally only for pairs of photons [1] and for pairs of protons [2]. In view of the importance of the nonlocality tested by Bell's inequality for our understanding of nature it is desirable to investigate more physical situations. In fact quite a number of proposals for other particles have been advanced [3,4]. Here I want to consider some proposals concerning atoms [5-9] where the atoms do not get entangled directly with one another, as in the other proposals concerning atoms [4], but indirectly, mediated through photons: two-level Rydberg atoms (upper state  $\state{e}$, lower state $\state{g}$) traverse a high-{\it Q} micromaser cavity one after the other, without spatial overlap of their wave functions (internal or translational). Each atom is initially in one of the Rydberg states. The cavity is in a number state $\state{n}$ in the resonant mode of the transition $\state{e} \leftrightarrow \state{g}$. After leaving the cavity the atoms pass a classical electromagnetic radiation field also in the resonant mode. Then each atom passes into a detector which registers the state  $\state{g}$  or  $\state{g}$, respectively. Within the cavity the first atom of a pair of successive atoms gets entangled with the cavity photons and emerges from the cavity in a superposition of the upper and the lower state. The second atom in traversing the cavity then becomes entangled with the modified cavity photons and therefore also with the first atom. Atom-atom correlations thus arise due to atom-photon correlations, i.e. due to correlations between non-identical particles. With this a new section of quantum mechanics is being considered.
\par
Moreover, such experiments are crucial for a realist interpretation in quantum mechanics which, in contrast to the traditional Copenhagen interpretation, predicts a Bell sum $S \! \le \! 2$  here. It holds that those Bell-type correlations which, according to the quantum mechanical formalism, arise between non-identical particles (different from particle-antiparticle pairs) are not to be taken seriously. This follows from a new conception of systems of several particles and of the situations where the particles become indistinguishable [10].
\par
The proposals by now appear feasible because the states of the Rydberg atoms can be detected with nearly 100 \% efficiency and because the techniques of treating high-{\it Q} microwave cavities are now well advanced. Thus, recently several papers have appeared [7-9] where the earlier proposals [5,6] are developed and experimental details are discussed.
\par
The features by which the proposed experiments [7-9] differ among themselves are the states of the atoms before they enter the cavity, the initial photon number state $\vert n \rangle$  of the cavity, its Rabi angle $\eta$ and the type of action of the classical field that is put behind the cavity. In the experimental realizations the two atoms of a pair of successive atoms may come from different sources and go into different detectors [7] or may come from one source and go into one detector [8,9].
\par
Specifically, in the present paper the following cases are  considered:
\par
   The atoms may initially be 
\par\noindent
(I)\    all in the lower state $\vert g \rangle$
\par\noindent
(II)\   all in the upper state $\vert e \rangle$ [8]
\par\noindent
(III) the first atom of a pair in state $\vert e \rangle$ and the second in $\vert g \rangle$  [7,9].
\par
   The classical field allows for two different freely adjustable Bell parameters to be chosen: the phase $\phi$ or the turning angle $\theta$. Specifically, the field may 
\par\noindent
(A) be chosen so that it acts as a $\pi /2$-pulse on the atoms, and when the first atom 
\par\  traverses the field the phase is set to $\phi_1$, and for the second to $\phi_2$ [8]
\par\noindent
(B) rotate the Bloch vector of the first atom through an angle 2$\cdot \theta_1$ and the second 
\par \ through 2$\cdot \theta_2$ [7,9].
\par
Finally, one may admit
\par\noindent
(i)  only one value of the cavity Rabi angle $\eta$  for all atoms that traverse the cavity
\par   [8], or
\par\noindent
(ii) different cavity Rabi angles $\eta_1$ and $\eta_2$     for the first and the second atom [7,9].

\par \bigskip
In this comment I want to compare the maximum possible values of the Bell sum {\it S}, calculated according to the standard formulas of quantum mechanics, in the 12 possible combinations of the cases (I), (II), (III) with (A), (B) and with (i), (ii). The cavity state $\vert n \rangle$ is left arbitrary. The published proposals mean the combinations [7]:\ (III B ii)\ {\it n}=0, [8]:\ (II A i), [9]:\ (III A ii)\ {\it n}=0. The motivation for this comparison is the following.
\par
{\it S}\  is calculated only in ideal situations, leaving out all effects of realistic circumstances which may also influence the value of {\it S}. That this influence can be quite considerable has been shown in [8] where the effects of energy dissipation in the cavity, the Poissonian arrival statistics of the atoms, and others were taken into account. It turned out that this reduced the maximum value of {\it S} by a factor of 0.6 (see Figs. 2, 3 and 5 of [8]). Realistic circumstances will always reduce the ideal value of {\it S} so that a priori an experimental situation should be chosen that gives an ideal value of {\it S} as high as possible in order to leave the realistic value still larger than 2. Of course in the final decision for a particular experimental setup the technical feasibilities will have the last word.
\par
The factor of 0.6 encountered in [8] would reduce the absolute maximum value of $S = 2 \sqrt2  = 2.83$ to $S = 1.70 < 2$ and thus make the experiment pointless. Nevertheless, determined experimental efforts in definite directions may still bring the realistic value nearer to the ideal one so that the experiment does make sense (see e.g. [7]).
\par\medskip\bigskip
\noindent
{\bf 2~~The quantum mechanical calculation}
\smallskip\par\noindent
Before the first atom has entered the cavity the state of the combined system consisting of the first atom (1) plus the second atom (2) plus the cavity is the product state
$$
\Psi_0 = \vert g_1 \rangle \vert g_2 \rangle \vert n \rangle \equiv \state{g_1, g_2, n} ,
\eqno(1)
$$
written down here for the case (I) where both atoms are initially in the lower state. Utilizing the fact that our two-level atoms can formally be treated like spin-\hbox{$1\over2$} particles the Hamilton operator of the resonant interaction of the {\it i}-th atom ($i=1,\,2$) with one cavity mode can be written as
$$
H_{\rm I}^{(i)} = \hbar \Omega_i (\sigma_+^{(i)} a + \sigma_-^{(i)} a^{\dag} ).
$$
The operators $ \sigma_+^{(i)} =$ $0\,1\choose0\,0$   and $\sigma_-^{(i)} =$ $0\,0\choose1\,0$  act on $\state{g_i}\equiv \hbox{$ 0 \choose 1 $}_i$, $\state{e_i}\equiv \hbox{$ 1 \choose 0 $}_i$, in matrix notation. The operators $a$ and $a^{\dag}$  annihilate and create the cavity photons in the resonant mode: $a \vert n \rangle = \sqrt n  \, \vert n-1 \rangle ,\, a^{\dag} \vert n \rangle = \sqrt{n-1}\, \vert n-1 \rangle.\, \,\Omega_i$ is the Rabi frequency and $\eta _i = \Omega_i t_i$ is the Rabi angle for the {\it i}-th atom with $t_i =$ time spent by the {\it i}-th atom in the cavity. Then straightforward operator algebra leads to a simple recipe for writing down the wave function of the system at different times: when the $i$-th atom traverses the cavity it causes the transition
$$
\state{e_i, n} \longrightarrow c_1^{(i)} \state{e_i, n} - {\rm i} s_1^{(i)} \state{g_i, n+1}
$$
$$
\state{g_i, n} \longrightarrow c_0^{(i)} \state{g_i, n} - {\rm i} s_0^{(i)} \state{e_i, n-1}
\eqno(2)
$$
where
$$
c_j^{(i)} = \cos(\eta_i \sqrt{n+j}\,), \quad s_j^{(i)} = \sin(\eta_i \sqrt{n+j}\,)
\eqno(3)
$$
and terms with negative photon numbers have to be omitted.
\par
Thus, the wave function $\Psi_1$ when the first atom has already left the cavity but the second has not yet entered it, in the three cases (I), (II), (III) is
$$
({\rm I}):\,\,\, \Psi_1 = c_0^{(1)}\state{g_1, g_2, n} - {\rm i}s_0^{(1)} \state{e_1, g_2, n-1}
$$
$$
({\rm II}):\, \Psi_1 = c_1^{(1)} \state{e_1, e_2, n} - {\rm i}s_1^{(1)} \state{g_1, e_2, n+1}
$$
$$
({\rm III}): \Psi_1 = c_1^{(1)} \state{e_1, g_2, n} - {\rm i}s_1^{(1)} \state{g_1, g_2, n+1} .
\eqno(4)
$$
Applying (2) in the analogous way to the second atom, the function $\Psi_2$  for the system when both atoms have traversed the cavity but not yet entered the classical field behind the cavity is
$$
({\rm I}):\,\, \Psi_2 = c_0^{(1)} c_0^{(2)} \state{g_1, g_2, n} - {\rm i} c_0^{(1)} s_0^{(2)} \state{g_1, e_2, n-1} - {\rm i} s_0^{(1)} c_{-1}^{(2)} \state{e_1, g_2, n-1}
$$
$$
- s_0^{(1)} s_{-1}^{(2)} \state{e_1, e_2, n-2} ,
$$
$$
({\rm II}):\, \Psi_2 = c_1^{(1)} c_1^{(2)} \state{e_1, e_2, n} - {\rm i} c_1^{(1)}s_1^{(2)} \state{e_1, g_2, n+1} - {\rm i} s_1^{(1)} c_2^{(2)} \state{g_1, e_2, n+1}
$$
$$- s_1^{(1)} s_2^{(2)} \state{g_1, g_2, n+2} ,
$$
$$
({\rm III}): \Psi_2 = c_1^{(1)} c_0^{(2)} \state{e_1, g_2, n} - {\rm i} c_1^{(1)} s_0^{(2)} \state{e_1, e_2, n-1} - {\rm i} s_1^{(1)} c_1^{(2)} \state{g_1, g_2, n+1} 
$$
$$
- s_1^{(1)} s_1^{(2)} \state{g_1, e_2, n} .
\eqno(5)
$$
In each of the functions $\Psi_2$ the first and the second atom have become entangled with each other due to the entanglement of each with the cavity photons.
\par
Now we consider the effect of the classical field together with the final detector.
\par
In case (A) the phase $\phi$ of the field is set to $\phi_1$, when the first atom traverses the field, and to $\phi_2$ for the second atom. The effect of the field on the two atoms leads to $\Psi_3 = U \Psi_2$   with $U = U^{(1)}U^{(2)}$ and
$$
U^{(i)} = \hbox{$ 1 \over \sqrt2 $} \left( \matrix{1 & -e^{{\rm i} \phi_i} \cr e^{- {\rm i} \phi_i} & 1 \cr} \right).
\eqno(6)
$$
The outcome of the final detector is assigned the value $+1$ if the atom is detected in the upper state  $\state{e}$, and $-1$ if it is in the lower state  $\state{g}$ . This is described by the operator $\sigma_z = $ $1\quad0\choose0\,\,\,-1$ acting on the atomic parts of the wave function $\Psi_3$. The average value $E$ of the product of the detector outcomes for atom 1 and atom 2, which will be needed in the Bell sum, can then be written as
$$
E = \langle \Psi_2 \vert L \vert \Psi_2 \rangle 
$$
where $L = L^{(1)}L^{(2)}$ and
$$
L^{(i)} = {U^{(i)}}^{\dag} \sigma_z^{(i)} U^{(i)} = \left(\matrix{0 & -e^{{\rm i} \phi_i} \cr -e^{-{\rm i} \phi_i} & 0 \cr} \right).
\eqno(7)
$$
This, with $\Psi_2$ of (5), yields
$$
E(\phi_1,\phi_2) = 2 \alpha \cos(\phi_2 - \phi_1)
\eqno(8)
$$
where $E = E_{({\rm IA})}, E_{({\rm IIA})}$  or $E_{({\rm IIIA})}$, respectively, and
$$
\alpha_{({\rm I})}= s_0^{(1)} c_0^{(1)} s_0^{(2)} c_{-1}^{(2)}
$$
$$
\alpha_{({\rm II})} = s_1^{(1)} c_1^{(1)} s_1^{(2)} c_2^{(2)}
$$
$$
\alpha_{({\rm III})} = - s_1^{(1)} c_1^{(1)} s_1^{(2)} c_0^{(2)}
\eqno(9)
$$
\par\noindent
in obvious notation.
\par
In case (B) the classical field turns the Bloch vector of the {\it i}-th atom through the angle $2\cdot \theta_i$ round the {\it y} axis, say. To describe this effect and that of the final detector we only have to replace the operator $L$ of case (A) by the `spin-projection' operator $O = O^{(1)} O^{(2)}$ with $\Psi_3 = O \Psi_2$ and
$$
O^{(i)} = \cos\theta_i \,\sigma_z^{(i)} + \sin\theta_i \, \sigma_x^{(i)} = \left( \matrix{\cos\theta_i & \sin\theta_i \cr \sin\theta_i & -\cos\theta_i \cr} \right) ,
\eqno(10)
$$
where $\sigma_x =$ $0\,1\choose1\,0$. This, with $\Psi_2$ of (5), yields
$$
E(\theta_1,\theta_2) = 2\alpha \sin\theta_1 \sin\theta_2 + 2 \beta \cos\theta_1 \cos\theta_2
\eqno(11)
$$
where $E = E_{({\rm IB})},\, E_{({\rm IIB})}$ or $E_{({\rm IIIB})}$, respectively. The $\alpha$s are the same as in (9), and
$$
\beta_{({\rm IB})} = \hbox{$1 \over 2$} \big [ \big (c_0^{(1)}c_0^{(2)}\big )^2 - \big (c_0^{(1)}s_0^{(2)}\big )^2 - \big (s_0^{(1)}c_{-1}^{(2)}\big )^2 + (s_0^{(1)}s_{-1}^{(2)})^2 \big ]
$$
$$
\beta_{({\rm IIB})} = \hbox{$1 \over 2$} \big [ \big (c_1^{(1)}c_1^{(2)}\big )^2 - c_1^{(1)}c_1^{(2)}s_1^{(2)}s_1^{(2)} -\big (s_1^{(1)}c_2^{(2)}\big )^2 + \big (s_1^{(1)}s_2^{(2)}\big )^2 \big]
$$
$$
\beta_{({\rm IIIB})} = \hbox{$1 \over 2$} \big[ -\big(c_0^{(2)}c_1^{(1)}\big)^2 + \big(c_1^{(1)}s_0^{(2)}\big)^2 + \big(s_1^{(1)}c_1^{(2)}\big)^2 - \big(s_1^{(1)}s_1^{(2)}\big)^2 \big]
\eqno(12)
$$
\par\medskip\bigskip\noindent
{\bf 3~~The Bell sum}
\smallskip\par\noindent
The considered Bell inequality, whose violation means nonlocality, is $S \le 2$ where $S$ is the Bell sum
$$
S = \big\vert  E(a_1 ,a_2 ) + E(a_1 ,a_2') + E(a_1',a_2 ) - E(a_1' ,a_2' ) \big\vert
\eqno(13)
$$
with $a_i  = \phi_i$  or $\theta_i$ in the cases (A) or (B), respectively.
\par
First we consider the cases (A), i.e. (I A), (II A), and (III A), where the adjustable Bell parameter is the phase $\phi$ of the classical field. The form (8) of the averages leads to the maximum Bell sum [11]
$$
S  = 4 \sqrt2 \vert \alpha \vert.
\eqno(14)
$$
So we have to find the maximum of $\vert \alpha \vert$ as a function of the Rabi angles $\eta_1$  and $\eta_2$ and the photon number $n$.
\par
First consider  the subcases (A i): $\eta_1 = \eta_2 = \eta$ . Then we have $s_0^{(1)} = s_0^{(2)}$ etc. in (9) and with (3) we get explicitly
$$
\alpha_{({\rm I})} = \sin^2(\eta \sqrt n \,) \cos(\eta \sqrt n \,) \cos(\eta \sqrt{n-1}\,) \quad n \ge 1
$$
$$
\alpha_{({\rm II})} = \sin^2(\eta \sqrt{n+1} \,) \cos(\eta \sqrt{n+1}\,) \cos(\eta \sqrt{n+2}\,) \quad n\ge 0
$$
$$
\alpha_{({\rm III})} = - \sin^2(\eta \sqrt{n+1}\,) \cos(\eta \sqrt{n+1}\,) \cos(\eta \sqrt n \,) \quad \ge 0 .
\eqno(15)
$$
Setting the last cosine equal to 1 and maximizing the rest one obtains $\vert \alpha \vert \le 2\sqrt3/9 = 0.385$ for all three $\alpha$s  and for all admissable $n$. This value is actually closely approached in many cases. E.g. in case (II A), $n=0$ one obtains $\alpha = 0.383$ for  $\eta =\pi/ \sqrt2$. The value $\vert \alpha \vert \le  0.385$ leads to $S \le 2.178$.
\par
Second consider the subcases (A ii): $\eta_1 \ne \eta_2$ . In these cases we have
$$
\alpha_{({\rm I})} = f(\eta_1,n) \sin(\eta_2\sqrt n \,) \cos(\eta_2 \sqrt{n-1}\,)
$$
$$
\alpha_{({\rm II})} = f(\eta_1,n+1) \sin(\eta_2 \sqrt{n+1}\,) \cos(\eta_2 \sqrt{n+2}\,)
$$
$$
\alpha_{({\rm III})} = f(\eta_1, n+1) \sin(\eta_2 \sqrt{n+1}\,) \cos(\eta_2 \sqrt n \,)
\eqno(16)
$$
where
$$
f(\eta,n) = \sin(\eta \sqrt n \,) \cos(\eta \sqrt n \,).
$$
The maximum of $f$ under variation of $\eta$  is 1/2 for any $n$. It remains to find the maximum of the $\sin \cdot \cos$ products in (16). These can never be exactly equal to 1 (except for $n=$ 0 and 1) but can come arbitrarily close to 1, as some insight and numerical studies show. As an example Fig. 1 shows that $\vert \sin(\eta_2 \sqrt 2 \,)\cos(\eta_2 \sqrt 1 \,) \vert > 0.98$ for $\eta_2 \approx $ 3.3 and 18.8. So we take $\vert \alpha \vert_{\rm max} = 1/2$ in the cases (16). According to (14) this results in
$$
S_{\rm max}  = 2.83 \approx 2 \sqrt 2 ,
\eqno(17)
$$
which is the absolute maximum of $S$.
\topinsert
\vskip 11 true cm
\centerline{Fig. 1. The function  $\sin(\eta_2 \sqrt2 \,) \cos(\eta_2)$  for  $0 \le \eta_2 \le 25$}
\bigskip
\endinsert

\par
Now we turn to the cases (B), i.e. (I B), (II B), and (III B) where the adjustable Bell parameter is the turning angle $\theta$ of the Bloch vector. The averages in these cases are given in (11), (12) and lead to
$$ 
S = 2 \big \vert \, \alpha [\sin\theta_1(\sin\theta_2 + \sin\theta_2') + \sin\theta_1' (\sin\theta_2 - \sin\theta_2') ]
$$
$$
\quad\quad + \beta [\cos\theta_1 (\cos\theta_2 + \cos\theta_2') + \cos\theta_1'(\cos\theta_2 - \cos\theta_2') ] \, \big \vert
\eqno(18)
$$
in all three cases. Here we follow the treatment in [7] and restrict ourselves to the special choice
$$
 \theta_1= 0,\quad \theta_1' = \pi/2,\quad \theta_2 = -
\theta_2'
\eqno(19)
$$
with the single free parameter $\theta_2$ . Maximizing $S$ with respect to $\theta_2$ then leads to
$$
S_{\rm max}  = 4 \sqrt{\alpha^2 + \beta^2} .
\eqno(20)
$$
Due to the special choice (19) we cannot be sure that the obtained maximum is always the true maximum. I suppose, however, that we come fairly close to it since (1) if $\alpha = \beta$ we have $S_{\rm max}  = 4 \sqrt2 \vert \alpha \vert$, which is the true maximum value for given $\vert \alpha \vert$, (2) in subcase (ii) ($\eta_1 \ne \eta_2$), treated below, the absolute maximum of $S$ = 2.83 is actually reached in all cases except one, (3) in subcase (i) ($\eta_1 = \eta_2 = \eta$) the fact that we never reach $S$ = 2.83 (see below) can be justified analytically: replacing $\Psi_2$ in (5) by the corresponding mixture can be shown to imply $\alpha$=0 in (11), i.e. $E_{\rm mixture}  = 2 \beta \cos\theta_1 \cos\theta_2$. From the definition of $E$ we have $\vert E \vert \le 1$, hence $\beta \le 1/2$. From (15) we obtained $\vert\alpha\vert \le 2\sqrt3/9$ in subcase (i), so with (20) we have the upper bound $S_{\rm max} \le 2.52$.
\par
Now, first consicer the subcases (B i): $\eta_1 = \eta_2 = \eta$. $S$ as a function of $\eta$ has been studied numerically and the largest values so obtained were
$$
S_{\rm (IB)max} = 2.33
$$
$$
S_{\rm (IIB)max}  = 2.33
$$
$$
S_{\rm (IIIB)max}  = 2.00
$$
for every considered value of $n$.
\par
Second consider the subcases (B ii): $\eta_1 \ne \eta_2$. The numerical studies yielded in these cases
$$
S_{\rm (IB)max} = S_{\rm (IIIB)max}  = 2.83
$$
for every considered value of $n$, and 
$$
S_{\rm (IIB)max} = 2.38 \; ({\rm for} \, n=0),\quad 2.32 \; (n=1),\quad 2.32 \;(n=2),\quad 2.41 \;(n=4).
$$

\topinsert
\vskip 11 true cm
\centerline{Fig. 2.  $S_{\rm max}$ in case (III B ii) as a function of $\eta_2$. $\eta_1=\pi/(4\sqrt2\,), n = 1$}
\bigskip
\endinsert

\par\bigskip

$$
\offinterlineskip \tabskip=0pt
\vbox{
\halign to 0.8\hsize
{\strut
\vrule #
\tabskip=0pt plus 100 pt
&\quad \hfil \rm# \hfil
&\vrule # &
&\quad\hfil# \hfil\quad
&\vrule#
&\quad\hfil # \hfil \quad
&\vrule #
&\quad \hfil # \hfil \quad
\tabskip=0pt
&\vrule #
\cr 
\noalign{\hrule}
& \hfil && (I) && (II) && (III) & \cr
& \hfil && (i) \quad (ii) && (i) \quad (ii) && (i)\quad  (ii) & \cr
\noalign{\hrule}
& (A) && 2.18 \quad 2.83 && 2.18 \quad 2.83 && 2.18 \quad 2.83 & \cr
& (B) && 2.33 \quad 2.83 && 2.33 \quad 2.41 && 2.00 \quad 2.83 & \cr
\noalign{\hrule}
}} 
$$
 
Table 1. Obtained maximum values of the quantum mechanical Bell sum $S$ in the cases (I): $\state{g_1, g_2}$, (II): $\state{e_1, e_2}$, (III): $\state{e_1, g_2}$, (A): phase $\phi$, (B): Bloch angle $\theta$,  (i): same cavity Rabi angle $\eta$, (ii): different cavity Rabi angles $\eta_1 \ne \eta_2$, as explained in the introduction.
\par\bigskip\bigskip

Table 1 summarizes the results. It is seen that when different cavity Rabi angles are admitted (experimentally realizable e.g. by different velocities of the two atoms of a pair), and only in these cases, the absolute maximum of $S$=2.83 can be reached. These are thus the most promising cases, so long as the experimental feasibilities of reaching the ideal value in the various cases are left out of account. Fig. 2 shows $S_{\rm max}$ in case (III B ii) as a function of $\eta_2$ for $0 \!\le\! \eta_2 \le 18.8$. It is seen that $S\! \ge \! 2$ roughly in the ranges $\eta_2 = 2.9$ -- 3.6,\   5.6 -- 5.9,\   9.5 -- 10.1,\   12.0 -- 12.7,\   18.5 -- 18.8, \   and that $S$ almost reaches 2.83 at $\eta_2 \approx$ 3.3, 12.3 and 18.8. 

\par\medskip\bigskip\noindent
{\bf References}
\par\noindent
\def\cit{\par\noindent\hangindent=16pt\hangafter=1}

\cit
[1] Kwiat, P. G., et al.: Phys. Rev. Letters {\bf 75}, 4337 (1995) and references therein
\cit
[2] Lamehi-Rachti, M. and Mittig, W.: Phys. Rev. D {\bf 14}, 2543 (1976)
\cit
[3] Home, D. and Selleri, F.: Riv. Nuovo Cimento {\bf 14}, 1 (1991); T\"ornqvist, N. A.: in Selleri, F. (ed.) {\it Quantum Mechanics Versus Local Realism} (New York, Plenum Press, 1988) p. 115; Home, D.: in Selleri, F. (ed.) {\it Quantum Mechanics Versus Local Realism} (New York, Plenum Press, 1988) p. 133;  Tixier, M. H. et al.: Phys. Lett. B {\bf 212}, 523 (1988); Datta, A. and Home, D.: Found. Phys. Lett. {\bf 4}, 165 (1991); Privitera, P.: Phys. Lett. B {\bf 275}, 172 (1992); Abel, S. A., Dittmar, M. and Dreiner, H.: Phys. Lett. B {\bf 280} (1992); Bertlmann, R. A. and Grimus, W.: {\it Quantum Mechanical Interference over Macroscopic Distances in the ${\rm B} \bar{\rm B}^0$ System}, Preprint University Wien, UWThPh-1996-57, October 3, 1996
\cit
[4] Livi, R.: Nuovo Cimento {\bf 48B}, 272 (1978); Lo, T. K. and Shimony, A.: Phys. Rev. A {\bf 23}, 3003 (1981); Fry, E. S., Walther, T. and Li, S.: Phys. Rev. A {\bf 52}, 4381 (1995)
\cit
[5] Oliver, B. J. and Stroud, C. R. Jr.: J. Opt. Soc. Am. B{\bf 4}, 1426 (1987)
\cit
[6] Phoenix, S. J. D. and Barnett, S. M.: J. Mod. Opt. {\bf 40}, 979 (1993); ibid. {\bf 40}, 1673 (1993)
\cit
[7] Freyberger, M. et al.: Phys. Rev. A {\bf 53}, 1232 (1996)
\cit
[8] L\"offler, M., Englert, B.-G. and Walther, H.: Appl. Phys. B {\bf 63}, 511 (1996). In this paper it is claimed that a violation of a Bell-type inequality is experimentally testable even under realistic circumstances. The Bell-type inequality used in this paper does not, however, test the same type of nonlocality as the original Bell inequality because the inequality in this paper is derived under the additional restriction on the hidden-variable theory that the expectation value of the product of the detector outcomes $E(\phi_1,\phi_2)$ in the case  $\phi_1=\phi_2=\phi$ is independent of $\phi$. In the particular physical situation under consideration quantum mechanics predicts $E(\phi,\phi) = E_0$ so that taking this as a restriction appears reasonable. In other physical situations, however, quantum mechanics leads to different predictions (see e.g. Barut, A. O. and Bo\v{z}i\'{c}, M.: Nuovo Cimento {\bf B 10}, 595 (1988)), and the general physical meaning of the restriction is not clear.
\cit
[9] Hagley, E. et al.: Phys. Rev. Lett. {\bf 79}, 1 (1997). This paper actually does not yet test the Bell inequality but only measures the production of correlated atom pairs. The authors state, however, that a test seems to be within reach of the described experimental setup.
\cit
[10] Jabs, A.: Brit. J. Philos. Sci. {\bf 54}, 405 (1992); Physics Essays {\bf 9}, 36, 354 (1996), quant-ph/9606017
\cit
[11] Bell, J. S.: {\it Speakable and unspeakable in quantum mechanics} (Cambridge University Press, Cambridge, 1987) p. 139

\bigskip

\bigskip
\centerline{\vrule width 5cm height 0.4 pt}
\bigskip
\bye